\documentclass[aps,twocolumn,amsfonts,amsmath,amssymb,prb]{revtex4}
\pdfoutput=1
\usepackage{graphicx}
\usepackage{graphics}
\newcommand*{\revision}[1]{{#1}}

\newcommand*{\revrem}[1]{}
\usepackage{amsfonts}
\usepackage{amssymb}
\usepackage{bm}

\usepackage{siunitx}

\begin{document}

\title{Capacitance of Nanoporous Carbon-Based Supercapacitors is a Trade-Off Between the Concentration and the Separability of the Ions}  
\author{Ryan Burt$^{1,*}$, Konrad Breitsprecher$^{2,*}$, Barbara Daffos$^{3,4}$, Pierre-Louis Taberna$^{3,4}$, Patrice Simon$^{3,4}$, Greg Birkett$^1$, X.S. Zhao$^1$, Christian Holm$^2$, Mathieu Salanne$^{4,5,6}$}
\affiliation{$^1$School of Chemical Engineering, University of Queensland, St Lucia, QLD 4072, Brisbane, Australia}
\affiliation{$^2$ICP, University of Stuttgart, Allmandring 3, 70569 Stuttgart, Germany}
\affiliation{$^3$CIRIMAT, Universit\'e de Toulouse, CNRS, INPT, UPS, 118 route de Narbonne, 31062 Toulouse Cedex 9, France}
\affiliation{$^4$R\'eseau sur le Stockage Electrochimique de l'Energie (RS2E), FR CNRS 3459, France}
\affiliation{$^5$Sorbonne Universit\'es, UPMC Univ Paris 06, CNRS, Laboratoire PHENIX, F-75005, Paris, France}
\affiliation{$^6$Maison de la Simulation, CEA, CNRS, Univ. Paris-Sud, UVSQ, Universit\'e Paris Saclay, F-91191 Gif-sur-Yvette, France}
\email{mathieu.salanne@upmc.fr}
\date{\today}

\begin{abstract}

Nanoporous carbon-based supercapacitors store electricity through adsorption of ions from the electrolyte  at the surface of the electrodes. Room temperature ionic liquids, which show the largest ion concentrations among \revision{organic} liquid electrolytes, should in principle yield larger capacitances. Here we show by using electrochemical measurements that the capacitance is not significantly affected when switching from a pure ionic liquid to a conventional organic electrolyte using the same ionic species. By performing additional molecular dynamics simulations, we interpret this result as an increasing difficulty of separating ions of opposite charges when they are more concentrated, i.e. in the absence of a solvent which screens the Coulombic interactions. The charging mechanism consistently changes with ion concentration, switching from counter-ion adsorption in the diluted organic electrolyte to ion exchange in the pure ionic liquid. Contrarily to the capacitance, in-pore diffusion coefficients largely depend on the composition, with a noticeable slowing of the dynamics in the pure ionic liquid.  


\end{abstract}

\maketitle







\label{sec:intro}

Supercapacitors are a promising electrochemical energy storage technology. \revision{They may be separated into two families of devices, depending on the electricity storage mechanism, i.e. the electric double-layer capacitors (EDLCs) and pseudocapacitors.~\cite{salanne2016a}. In EDLCs, the charge is stored by a simple process of ion adsorption at the surface of the electrodes}, even if at the microscopic scale complex mechanisms are at play. The most prominently employed electrode materials are porous carbons with high accessible surface area. Research and development of such EDLCs is being driven by demand from emerging applications which require efficient energy storage, notably renewable energy,\cite{Gogotsi2011} electric vehicles,\cite{Faggioli1999} and smart-grid management.\cite{Dunn2011,Simon2008a}


Currently, commercial supercapacitors have energy densities on the order of \SI{5}{\watt\hour\per\kilogram}.\cite{Simon2008a}  There is a large ongoing interest in the design of electrolytes with improved performance over the traditional organic liquids, such as acetonitrile-based electrolytes. Room temperature ionic liquids (RTILs) have raised a lot of interest due to their high  window of electrochemical stability,\cite{Rogers2009} high thermal stability,\cite{Huddleston2001} and low toxicity.\cite{Buzzeo2004} RTILs are also highly tunable, and there are many possible combinations available which is beneficial as they can be selected to match specific electrodes.~\cite{Largeot2008} However, when used as a pure electrolyte in EDLCs with highly microporous carbon electrodes, they suffer from high viscosity and poor conductivity.\cite{Guerfi2010} Superconcentrated electrolytes, which are highly-concentrated mixtures of RTILs and organic solvents, may offer a good compromise for these target properties, while keeping most of the RTIL benefits. They are therefore considered as promising alternatives in EDLCs.
%

A difficulty for choosing the optimal electrolyte is that the ions they contain  adsorb inside electrified nanoporous carbon electrodes, in which the structure changes significantly due to strong electric field and confinement effects. Although the molecular mechanisms that contribute to enhanced capacitance have been partially resolved,~\cite{salanne2016a} no generic theory is available yet. Indeed, elucidating this knowledge by traditional experiments is challenging due to the difficulty of probing inside three dimensional nanoporous structures. Complex \textit{in situ} techniques have to be employed.~\cite{Richey2012,Banuelos2013,Forse2015} Mean-field theories and molecular modeling have also been used to complement such experiments.~\cite{Fedorov2014,Burt2014,Feng2013a,Merlet2013f,Vatamanu2013,he2015aa} Despite the increasing activity on this topic, only a limited amount of systems have been investigated so far. In particular, the difference between RTILs and organic electrolytes with similar ions has mostly been studied for the simple case of flat, non-porous graphite electrodes. Using molecular dynamics (MD) simulations of mixture of RTILs and acetonitrile (ACN), Feng et al.\cite{Feng2011b} showed that the concentration of ACN only results in a small change in capacitance. Larger variations have recently been observed by Uralcan {\it et al.}\cite{Uralcan2016}. \revision{By using a simple analytical model, Lee and Perkin showed that mixing ionic liquids with solvents is also an efficient way to avoid hysteresis effects due to phase transitions in the adsorbed layer.}~\cite{lee2016b}

 In this work we study the behavior of electrolytes over concentrations ranging from dilute to \revision{the maximum concentration} (\revision{reached at} the pure RTIL) in nanoporous carbon electrodes. Cyclic voltammetry  experiments show that the capacitance does not vary substantially with the ionic concentration. By using molecular dynamics (MD) simulations on similar systems we elucidate the molecular mechanisms leading to this behavior. 




\label{sec:results}


\begin{figure}[ht]
	\centering
	\includegraphics[width=\columnwidth]{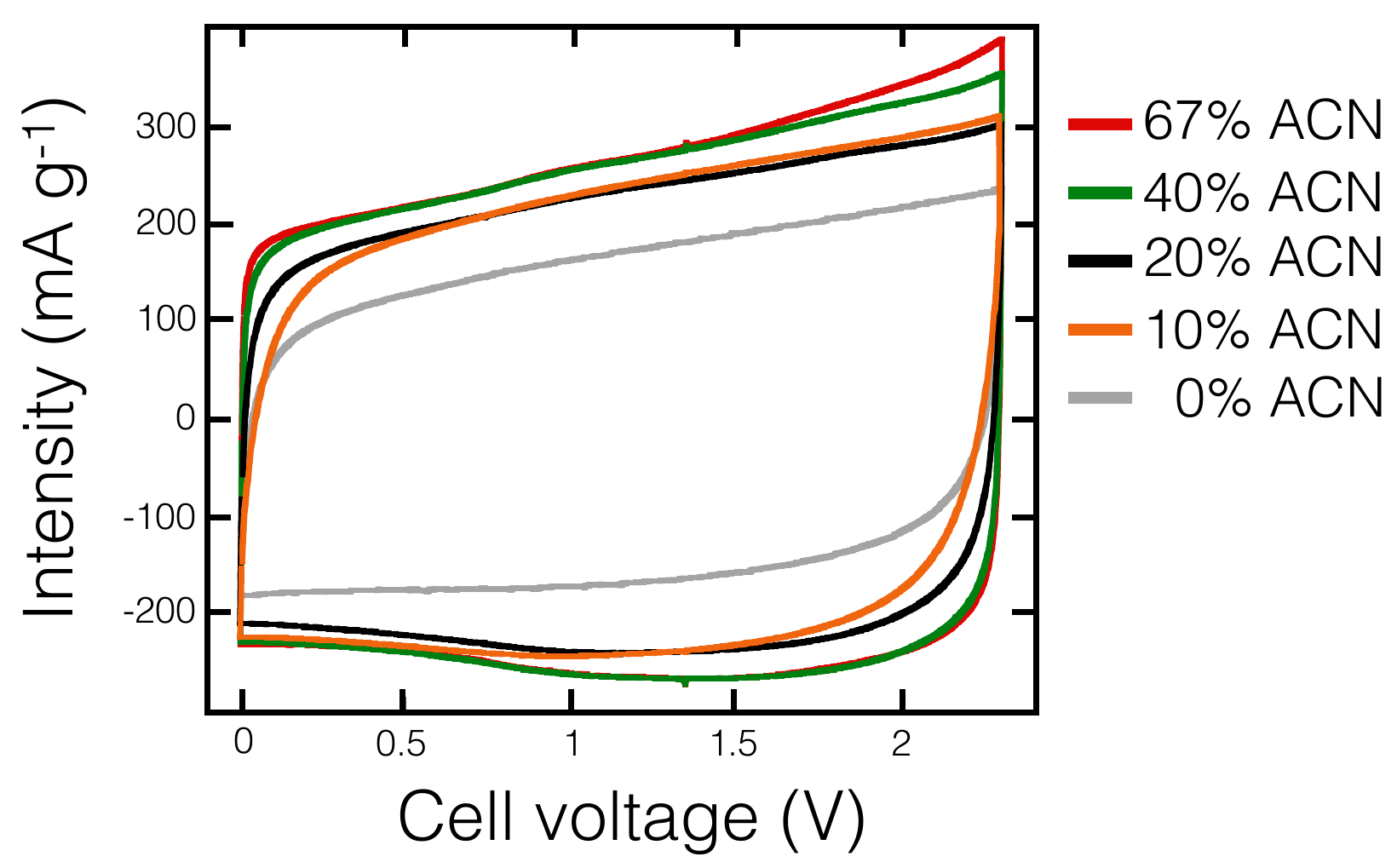}
	\caption{Cyclic voltammograms of supercapacitor cells assembled with two electrodes based on carbide-derived carbons (CDC), in electrolytes composed of EMIM-BF$_4$ mixed with ACN at several mass fractions. The potential scan rate is 5~mV~s$^{-1}$.}
	\label{fig:CV}
\end{figure}

\noindent
We used carbide-derived carbon (CDC) electrodes prepared by chlorination of TiC powder at 800~$^\circ$C, which are typical nanoporous carbons with a narrow pore size distribution -- here with pore sizes ranging  up to 1.5~nm, and a mean diameter \revision{located between 0.66~nm and 0.77~nm (as given by NLDFT and QSDFT, respectively, see the Methods section)}. The electrolyte is composed of a typical ionic liquid, 1-ethyl-3-methylimidazolium tetrafluoroborate (EMIM-BF$_4$), mixed with ACN at mass fractions ranging from 0~\% to 67~\% ACN.  Cyclic voltammetry (CV) experiments were carried out for potentials ranging between 0 and 2.3~V, at a scan rate of 5~mV~s$^{-1}$ (see the Methods section). The recorded voltammograms are shown in Figure \ref{fig:CV}. The CVs look very similar across the whole range of composition, except for the pure RTIL. The corresponding specific capacitances are provided in Table \ref{tab:Capacitances}. 

\begin{table*}[ht]
	\centering
	\caption{Summary of the experimental and simulated capacitances. The experimental data is obtained from CV by integration of the electric current during the discharge of the cell. The MD simulation data is extracted from the charge accumulated at the surface of the electrodes for an applied voltage of 1~V.\label{tab:Capacitances}}
	\begin{tabular}{|l|l|l|l|l|l|}
		\hline
		ACN mass \%        & 0     & 10   & 20   & 40   & 67   \\ \hline
		Ion conc. (mol L$^{-1}$ at $T$~=~298~K) & 6.40 & 5.28 & 4.58 & 3.01 & 1.51 \\ \hline
                $C$ (F g$^{-1}$) / CV, $T$~=~298~K, scan rate = 5 mV s$^{-1}$ & 70 & 100 & 100 & 105 & 105 \\ \hline
                \revision{$C$ (F g$^{-1}$) / CV, $T$~=~298~K, scan rate = 1 mV s$^{-1}$} & \revision{80} & -- & -- & -- & -- \\ \hline
                $C$ (F g$^{-1}$) / CV, $T$~=~373~K, scan rate = 5 mV s$^{-1}$ & 150 & -- & -- & -- & -- \\ \hline
                $C$ (F g$^{-1}$) / MD, $T$~=~340~K & 140 & 145 & 145 & 125 & 115 \\ \hline
	\end{tabular}
	
\end{table*}

\begin{figure*}[ht]
	\centering	
	\includegraphics[width=.9\textwidth]{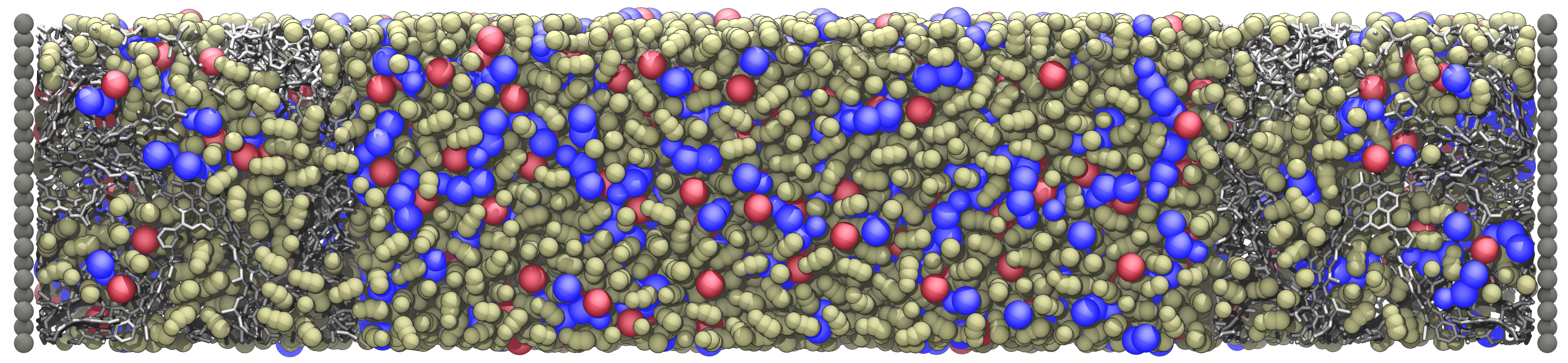}
	\caption{\hangafter=5
		Simulation cell of an EMIM-BF$_4$ and ACN electrolyte mixture surrounded by disordered porous Ti-CDC800 electrodes. Color 
scheme; blue: 3-site EMIM$^+$ molecules, red: single-site BF$_4^-$ molecules, pale yellow: 3-site 
ACN molecules, silver: carbon electrode atoms. Gray molecules cap the cell in the $z$-dimension, as this is non-periodic.} 
	\label{fig:Snapshot}
\end{figure*}

 In order to interpret these experimental results, we  performed MD simulations on similar systems. We employed a method in which the electric potential between the electrodes is constrained to a specified value,~\cite{Reed2007,Pounds2009} which was chosen to be 1~V here. This is enabled by calculating the individual charge on each carbon atom within the electrode at every time-step of the simulation, so that the specified potential is reached. The distribution of charge is determined and induced by the location of charged electrolyte species, and thus can fluctuate with the movement of liquid molecules. This is commonly referred to as constant potential MD. \revision{Another difficulty is to account for the shape of the porous carbon and to the pore size dispersion in simulations\cite{Kondrat2012}, so that the realistic CDC carbon structures of Palmer {\it et al.}\cite{Palmer2010} were employed for the electrodes.} A typical snapshot of the simulated system is shown in Figure \ref{fig:Snapshot}, and more technical details about the simulations are also provided in the Methods section. 

From the simulations, the integral capacitance is easily determined from the average of the total charge accumulated on the electrode. The simulated values at $T~=~340$~K are also provided in Figure \ref{tab:Capacitances}. Unfortunately, it was not possible to perform the simulations at room temperature (298~K) because the dynamics of the pure RTIL is very slow, making equilibration times too long.   Overall, there is a qualitative agreement between simulations and experiments, although the simulated capacitances overestimate the experimental results. The only large discrepancy corresponds to the pure RTIL. \revision{The comparison is improved when a slower scan rate of 1~mV~s$^{-1}$ is used for measuring the experimental capacitance (80~F~g$^{-1}$ instead of 70~F~g$^{-1}$), which points towards a kinetic cause for this discrepancy.} In order to confirm the origin of this difference, an additional experiment were performed for this system at a greater temperature (373~K), yielding a two-fold increase of the capacitance. Temperature effects on the capacitance are expected to be \revision{small},~\cite{Vatamanu2014} \revision{unless the temperature at which the experiments are carried is too low to allow for a good diffusivity of the ions~\cite{Feng2013a}, in which case the capacitance measurement is hampered by a large ohmic drop -- this is the case here}. For the other points, the overestimation of the capacitance in simulations is likely due to the simplifications introduced by the model (mainly the coarse-graining of the molecules and the absence of field penetration effects in our electrode model, which treats carbon as an ideal metallic conductor).~\cite{Kornyshev2014} It is consistent with our previous work using 1-butyl-3-methylimidazolium hexafluorophosphate as the electrolyte and CDC electrodes with slightly larger pores.~\cite{pean2015b}

The main conclusion we can draw from  Table \ref{tab:Capacitances} is that the capacitance does not vary much with concentration, at most by 5~\% for the experiments (if we discard the pure RTIL), and by 25~\% for the simulations, despite a fourfold change between the two extreme cases we considered. \revision{This result contrasts with the more dilute regime (100 mmol~L$^{-1}$ or less, i.e. not at all in the supercapacitor regime, for which large variations of the capacitance are observed, especially for low applied potentials).} It means that increasing the concentration of the electrolyte is not an efficient way to increase the capacitance, which can be due to several factors, that we try to examine below.

\begin{figure*}[ht]
		\includegraphics[width=0.45\textwidth]{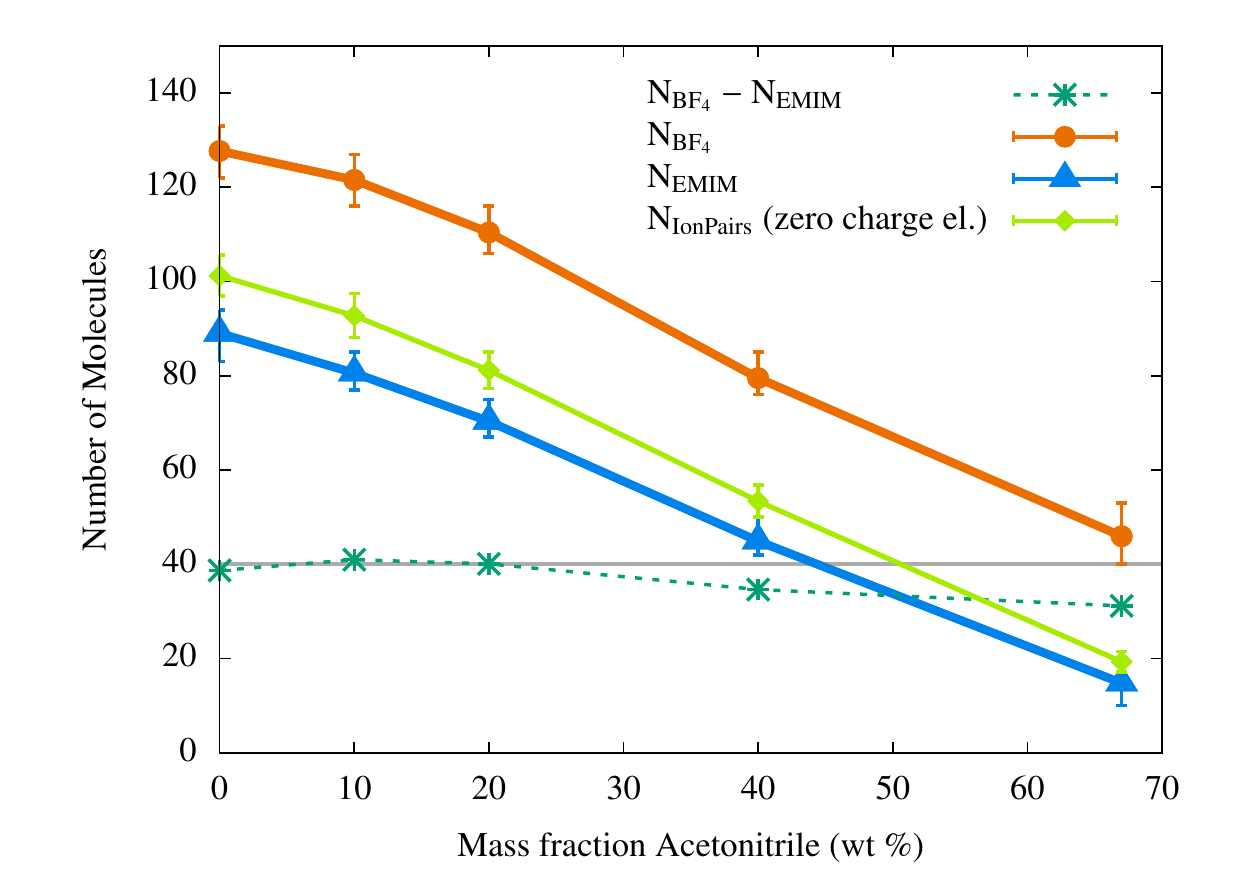}
		\includegraphics[width=0.45\textwidth]{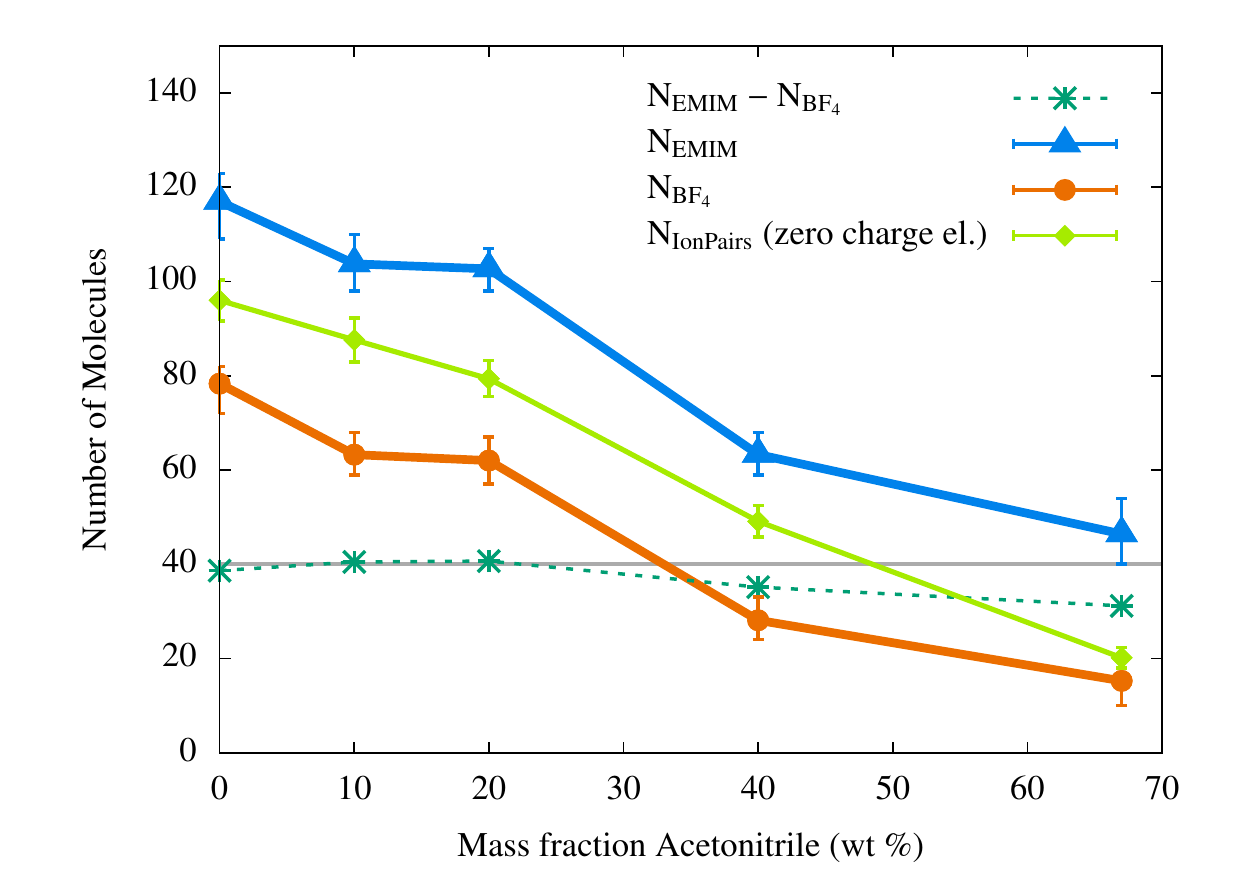}
	\caption{Electrolyte composition inside the electrodes (left: positive electrode, right: negative electrode). Light green: uncharged CDC electrodes; Orange/Blue: electrodes held at a constant potential difference of 1~V. Dark green: difference between the counter-ions and the co-ions numbers for the 1~V simulation.}
	\label{fig:partcounts}
\end{figure*} 



A simple hypothesis is that the number of ions adsorbed inside the nanoporous carbons is similar for all the systems, i.e. that a saturated regime is quickly reached in which the pores are totally filled with ions. The MD simulations provide the electrolyte compositions inside the pores as a function of ACN concentration, which are shown in Figure \ref{fig:partcounts} for uncharged electrodes and for an applied voltage of 1~V between them. A large linear variation is observed in all cases, ruling out the existence of such a saturated regime. 
Instead, we observe that the difference between the number of counter- and co-ions, hence the net charge (dashed green line), is almost constant at 1~V, which results in similar total induced charge in each electrode despite the large variation in the total number of ions.  When applying a voltage, some differences between the electrodes become apparent. \revision{More ions are adsorbed inside the positive electrode,} an effect which can be traced back to the smaller size of BF$_4^-$ anions with respect to the EMIM$^+$ cations. In the positive electrode the number of both counter- and co-ions decreases smoothly with ACN mass fraction. However, in the negative electrode, the decrease is less visible, with an almost negligible change in ion number when increasing the  ACN mass fraction from 10 to \SI{20}{\percent}.  At the highest ACN mass fraction, there is very little difference in electrolyte composition inside both electrodes, which is to be expected as the differences due to ion size then plays a minor role. 

\begin{figure}[ht]
	\centering
	\includegraphics[width=\columnwidth]{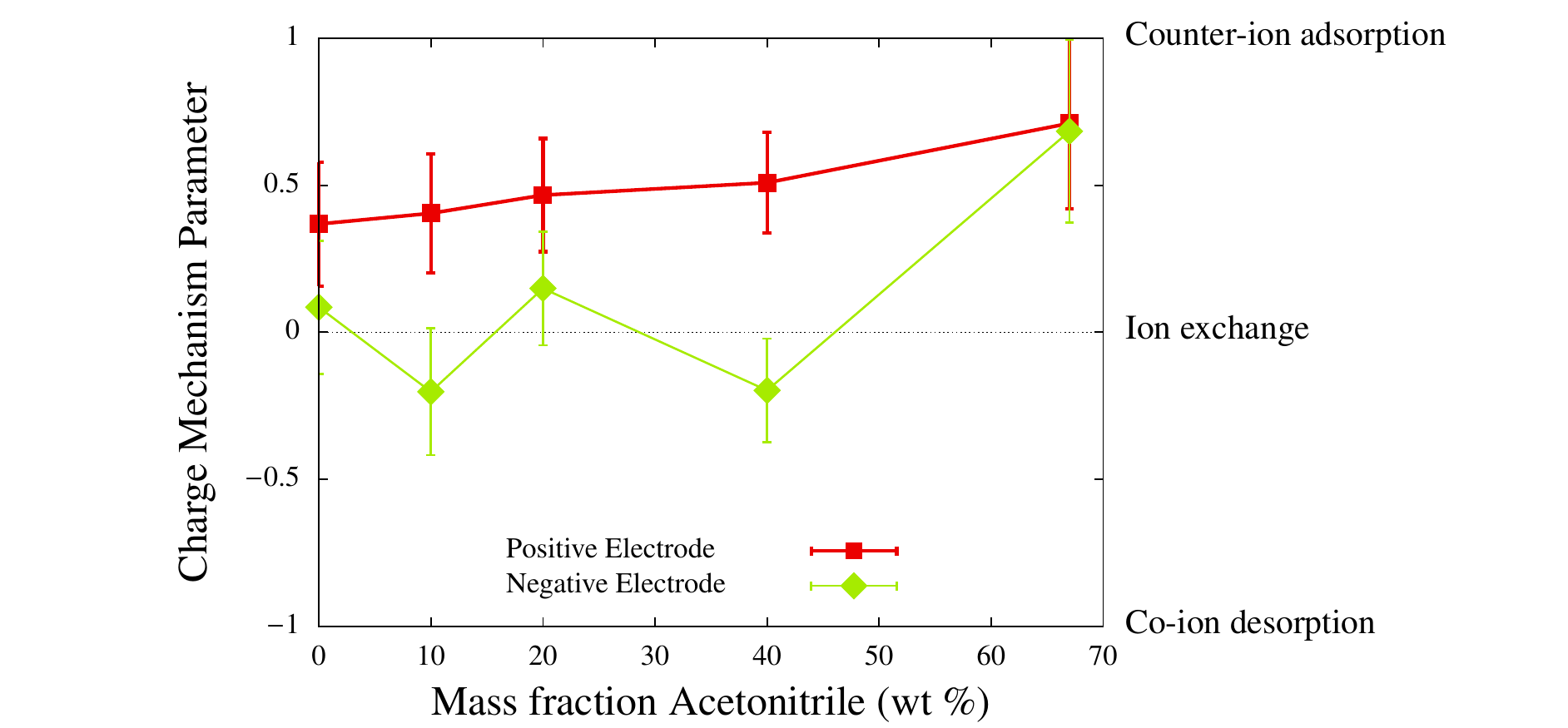}
	\caption{Variation of the charging mechanism parameter with the ACN mass fraction for the positive and negative electrodes at 1 V total potential.}
	\label{fig:ads_param}
\end{figure}

In order to understand better the influence of ionic concentration on the charge storage mechanism, and the differences between the positive and negative electrodes, we use a charging mechanism parameter $X$ recently introduced by Forse et al.,\cite{Forse2016} which quantifies the origin of charge induction between three different mechanisms: counter-ion adsorption, ion exchange, or co-ion desorption. It is defined for each electrode as
\begin{equation}
X=\frac{N(\Psi)-N(0{\rm V})}{\left[ \mid Q(\Psi)\mid - \mid Q(0~V)\mid\right]/e}
\end{equation}
\noindent where $N(\Psi)$ is the total number of in-pore ions at a given voltage $\Psi$, $Q(\Psi)$ is the corresponding electrode charge and $e$ is the elementary charge. $X$ is plotted in Figure \ref{fig:ads_param} as a function of the ACN mass fraction. This parameter takes a value of 1 for counter-ion adsorption, 0 for ion exchange and -1 for ion desorption; while intermediate values point to a combination of two such mechanisms. We observe clear trends for both electrodes. Ion exchange dominates in the pure RTIL ($X \approx$~0.1 and 0.3 in the negative and positive electrodes respectively). In the positive electrode the mechanism progressively switches to counter-ion adsorption, while in the negative one ion exchange dominates until the most diluted concentration is reached. In both cases a value of 0.7 is obtained for $X$ in the 1.5~M solution (ACN mass fraction 67~\%). The evolution of the mechanisms gives a first hint why the RTIL shows a low capacitance despite the high ionic concentration. In fact, at 0~V, the pores are completely filled of ions already, and adsorbing a counter-ion therefore requires the simultaneous desorption of a co-ion. The strong Coulombic attraction between ions of opposite charge \revision{therefore impedes an efficient charching in this system} (even if this interaction is screened in the nanopores, leading to a superionic state as was shown by Kondrat and Kornyshev).~\cite{Kondrat2011c} In the solvent-based system, the ions are solvated and they interact less intensely, such that this effect becomes less important.

\begin{figure*}[ht]
		\includegraphics[width=0.45\textwidth]{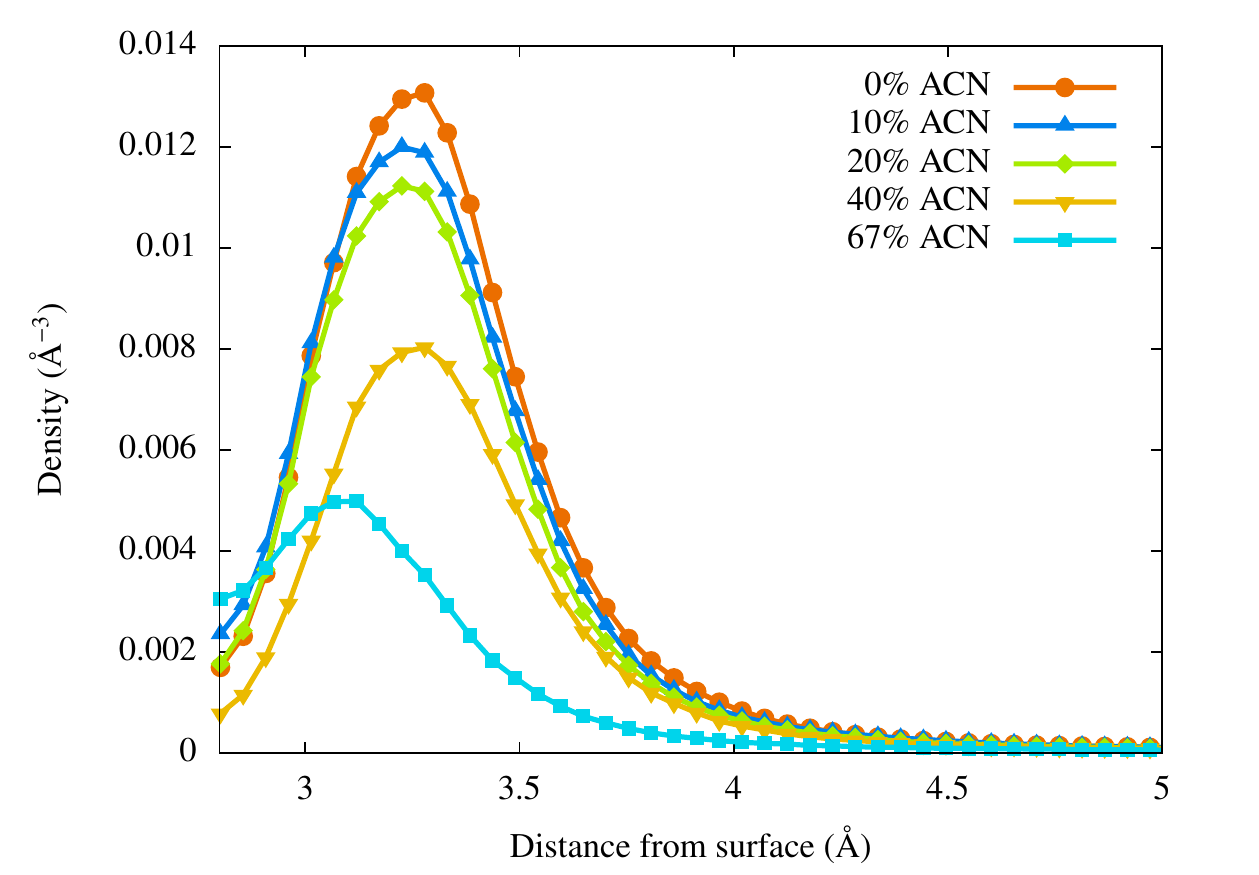}
		\includegraphics[width=0.45\textwidth]{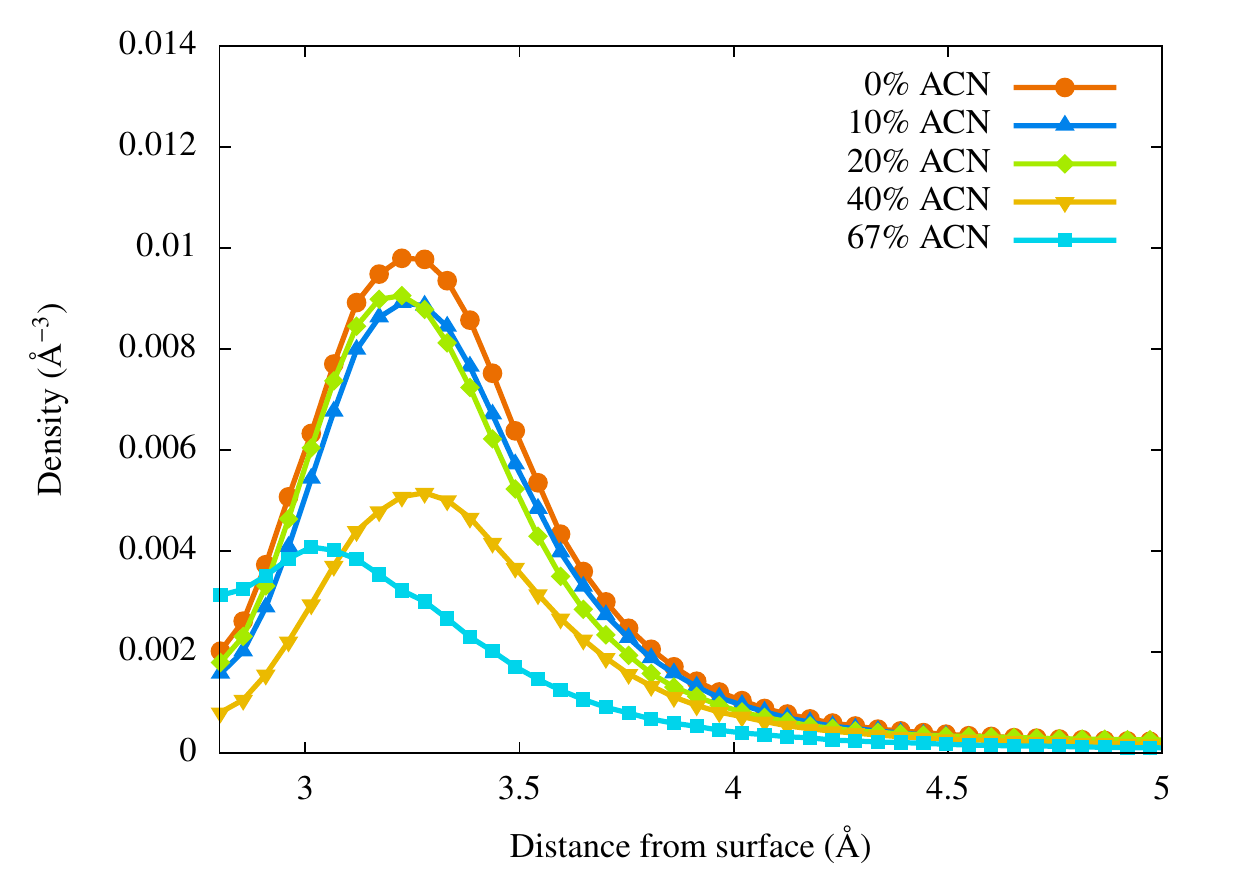}
	\caption{Counter-ion distance from the internal surface of the electrodes at 1 V total potential (left: positive electrode, right: negative electrode).}
	\label{fig:counter-ion_densities}
\end{figure*}


\noindent


However, solvation effects do not explain why the ACN-based system show a storage mechanism dominated by counter-ion adsorption. Figure \ref{fig:counter-ion_densities} shows the density of counter-ions with respect to the distance from the internal surface of the positive and negative electrodes for different fractions of ACN in the electrolyte. The position of the internal surface is determined as in our previous work,\cite{Merlet2012e} firstly by probing the carbon electrode atoms outwards with an argon atom to determine which volumetric sections of the electrode are accessible to the electrolyte molecules. The distance of the molecules inside this volume to the nearest carbon atom can then be calculated, and the median position of the ions with respect to the surface can be determined. From Figure \ref{fig:counter-ion_densities}, we see  that there is no change in the position of the density peaks in both electrodes, except for the most dilute electrolyte. However, the magnitude of the peak decreases consistently with the ionic concentrations shown in Figure \ref{fig:partcounts}. This means that the access to the surface becomes facilitated for counter-ions, which then have to replace solvent molecules that do not interact strongly with the surface or other adsorbed counter-ions. Another interesting observation of Figure \ref{fig:counter-ion_densities} is that the position of the peak remains constant until the ACN fraction is increased to 67 \% mass, where the peak shifts suddenly approximately to \SI{0.2}{\angstrom} closer to the surface. There is no clear explanation for this phenomena, but it is possible that the smaller ionic density allows the ions to pack more efficiently. \revision{As for the solvation numbers of the ions, their variation is linear, with no particular change at a given concentration.}


\begin{figure*}[ht]
	  \includegraphics[width=0.45\textwidth]{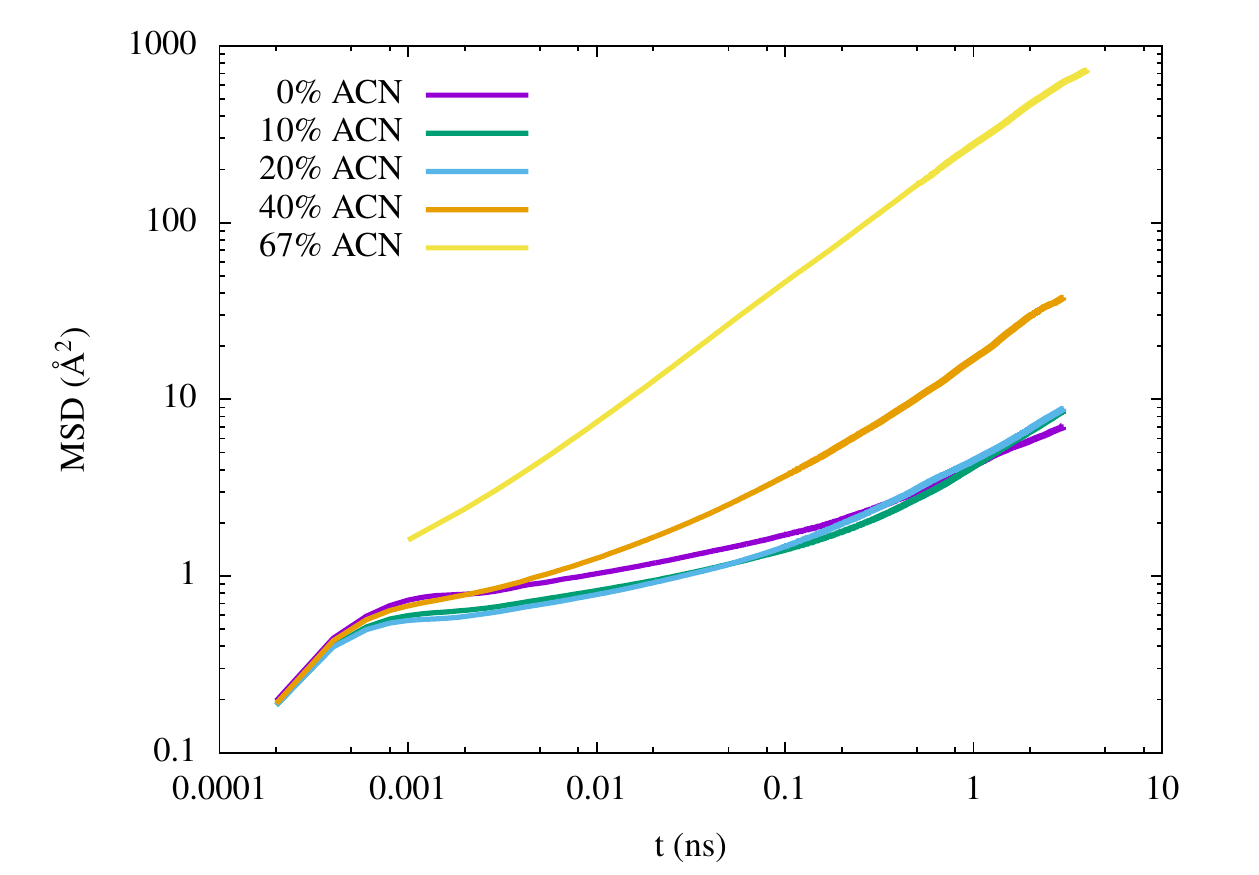}
	  \includegraphics[width=0.45\textwidth]{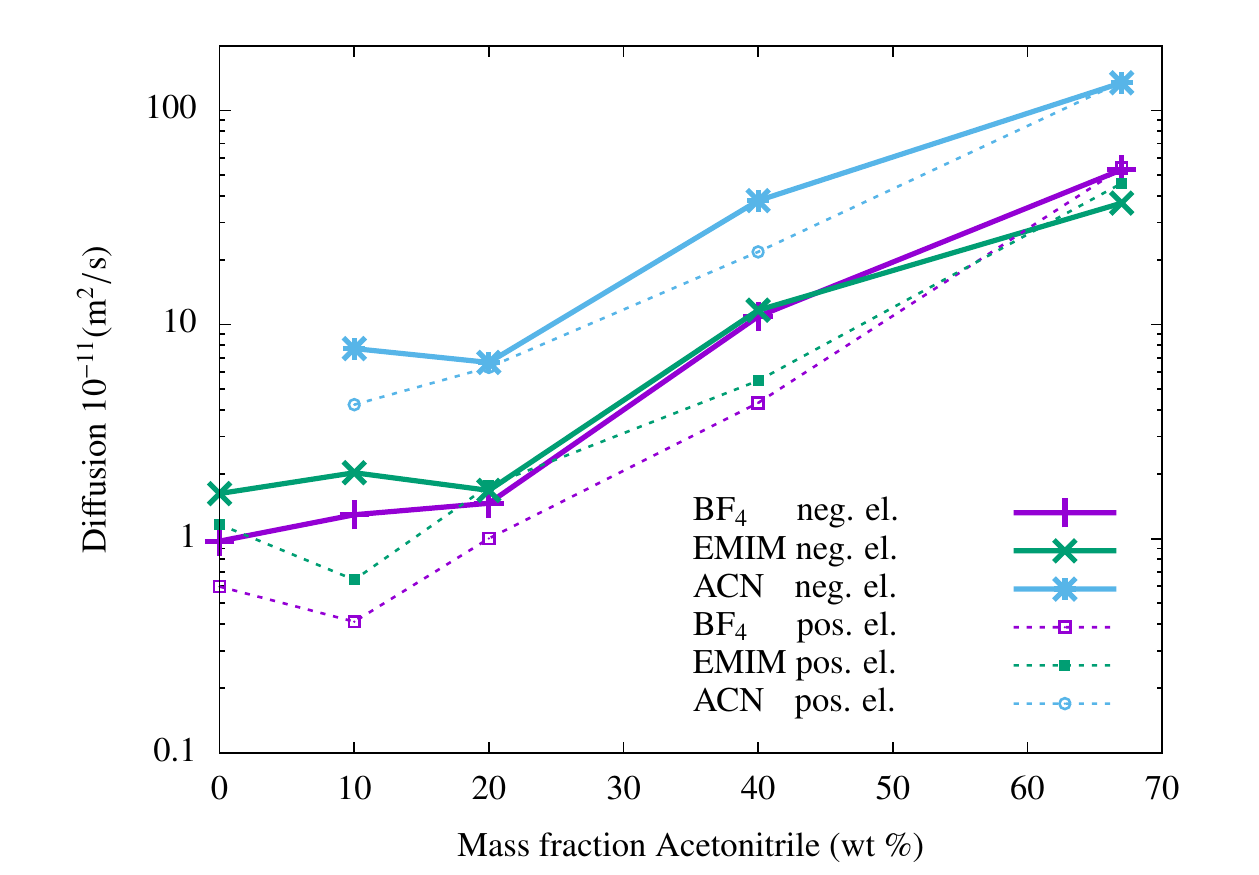}
\caption{\hangafter=5
	Mean squared displacement of BF$_4^-$ counter-ions inside the positive electrode (a), and diffusion coefficients of all molecular species inside both electrodes (b)}
\label{fig:msd_and_diffusion_pos}
\end{figure*}

Finally, one of the benefits of adding ACN to the electrolyte is to increase the mobility of ions in the bulk so that they diffuse more rapidly  into and out of the pores, which increases the capacitive performance at greater current densities. This can be clearly deduced from the mean squared displacements of the various species. As shown in Figure \ref{fig:msd_and_diffusion_pos}a for the BF$_4^-$ anions, a caging regime is observed for the pure RTIL, and the diffusive regime is reached for times greater than 1~ns. When the concentration of ACN increases, the extent of this caging regime decreases, and it completely vanishes for the most diluted system.  In  Figure \ref{fig:msd_and_diffusion_pos}b the extracted self-diffusion coefficients of all molecular species are given for each electrode. The solvent molecules diffuse around 10 times faster than the ions, and the EMIM$^+$ cations are slightly more mobile than the smaller BF$_4^-$ anions, which is consistent with previous results.\cite{Merlet2012g} It can be seen that significant gains in diffusion are not reached before the ACN mass fraction is 40~\%. Below this value, the concentration of ACN appears to be too low to decrease the ion-ion correlations enough to facilitate increased ion mobility.



\label{sec:conclusions}

In conclusion, in this work we have performed constant potential CV experiments and MD simulations to determine the influence of ACN fraction in an RTIL based electrolyte on capacitance and ionic structure inside nanoporous CDC electrodes \revision{(in the regime of concentrations which are relevant for EDLCs)}. Qualitative agreement between experiment and simulation was found, with an overall overestimation of the capacitance by simulations due to temperature effects and to the absence of some phenomena because of several necessary simplifications introduced in the models (coarse-graining of the electrolyte, electrode treated as an ideal conductor). Both methods show a small fluctuation of the capacitance despite the drastic variation of ionic concentration (with a factor of 4 between the most dilute and the most concentrated situations). 
The simulations show noticeable changes in the charging mechanisms with the ACN concentration. Indeed, the pure RTIL is characterized by a large ionic density close to the carbon surface. Adding a counter-ion upon application of a voltage can thus be achieved by exchanging a co-ion only, which requires to break the cation-anion associations which are favored by the strong Coulombic interactions (albeit the latter are screened by the electrostatic potential due to the metallic walls). In an ACN-based system, it is not necessary to remove co-ions for adsorbing additional counter-ions, resulting  in a mechanism dominated by counter-ion adsorption. The transition from one mechanism to another is smoother in the positive electrode than in the negative one. In parallel, the dynamics of the ions increases by one order of magnitude when switching from the pure RTIL to the diluted electrolyte.

Since the ionic concentration  does not affect much the capacitance, the choice of electrolyte for efficient EDLCs should therefore be made by comparing the ionic conductivity, which determines the internal resistance of the device, and the operating voltage it allows. However, these two quantities vary in opposite ways in concentrated electrolytes,~\cite{beguin2014a} so that the optimum will necessarily be a trade-off between these quantities.~\cite{krause2011a,pohlmann2015a} \revision{Other limitations such as the operating temperature or the long-term stability may also be taken into account.} It is worth noting that several ionic liquids may be combined,~\cite{lin2011a} which increases further the number of potential electrolytes. Computational screening approaches have therefore recently been proposed to tackle this difficult problem.~\cite{schutter2015a} 


\section{Methods}
\label{sec:methods}

\subsection{Electrochemistry experiments}

\begin{table}[ht]
	\centering
	\caption{Electrolyte compositions}
	\label{tab:partNum}
	\begin{tabular}{|l|l|l|l|l|l|}
		\hline
		ACN mass \%        & 0 (pure IL)    & 10   & 20   & 40   & 67   \\ \hline
		Ion conc. (mol/L at 298.15 K) & 6.40 & 5.28 & 4.58 & 3.01 & 1.51 \\ \hline
		Ion pairs                & 600  & 601  & 608  & 326  & 324  \\ \hline
		ACN molecules                 & 0    & 322  & 733  & 1048 & 3172 \\ \hline
	\end{tabular}
\end{table}

EMIM-BF$_4$ ionic liquid (Solvionic, France) and ACN (Acros organic, France) were used as purchased. They were mixed at room temperature, yielding 5 different ionic concentrations as listed in \ref{tab:partNum}. Carbide-derived carbon (CDC) powder (Y-Carbon, USA) was prepared by chlorination of TiC powder at 800~$^\circ$C  as reported elsewhere~\cite{chmiola2006a,Largeot2008}. Pore Size Distribution (PSD) is difficult to measure precisely in the case of nanoporous carbons~\cite{salanne2016a}. Here it was obtained from Ar-isotherms using NLDFT \revision{and QSDFT} models (NOVAe SERIES software, QUANTACHROME, USA). \revision{The measured average pore size were respectively of 0.77 and 0.66~nm. } Active films were made by mixing 95~wt\% CDC with 5~wt\% polyTetraFluoroEthylene (PTFE) binder. Once calendered, 8~mm diameter electrodes electrodes were cut. The active film thickness was around 300~$\mu$m, with a weight loading of 15~mg~cm$^{-2}$. Active films were laminated onto treated aluminum current collectors~\cite{Portet2005} and two layers of 25 $\mu$m-thick porous PTFE were used as a separator. A silver wire was used as a pseudo-reference electrode, for monitoring the negative and positive electrode potentials, separately during the cell cycling. Cell assembly was done in a glove box under an argon atmosphere ($<$ 1 ppm of O$_2$ and H$_2$O content) in 3 electrode Swagelok cells. Cyclic voltammetry experiments were carried out between 0 and 2.3~V, at a scan rate of 5~mV~s$^{-1}$.

\subsection{Molecular dynamics simulations}

Electrolyte compositions used can be found in \ref{tab:partNum}. Coarse-grained models for ACN and EMIMBF$_4$ were obtained from Edwards et al.\cite{Edwards1984} and Merlet et al.\cite{Merlet2012g} respectively. The simulation boxes consist of EMIMBF$_4$ and ACN surrounded by two symmetric carbide-derived carbon electrodes, with an average pore size of approximately 0.75~nm. The model electrodes were obtained by molecular dynamics from Palmer et al.,\cite{Palmer2010} \revision{by quenching a sample of liquid carbon consisting of approximately 4000 atoms with a rate of 20~$\times$~10$^{12}$~K~s$^{-1}$}. Finally the carbon Lennard-Jones parameters of $\sigma$ = 0.337~nm and $\epsilon$ = 0.23~kJ~mol$^{-1}$ were obtained from Cole et al. \cite{cole1983}

Initial equilibration and pore filling was performed in open source MD package GROMACS.\cite{hess08a} Firstly, the liquid electrolytes were equilibrated at \SI{340}{\kelvin}. The liquid boxes were then placed between the electrodes and allowed to relax in the NVT ensemble. Pore filling rate was enhanced by cycling the constant charge values of the electrode atoms between $\pm$ 0.01 e for 10 iterations. The boxes were then equilibrated at zero electrode charge for \SI{500}{\pico\second}. Second stage equilibration was performed with an in-house code. The distance between the electrodes was tuned by position rescaling for each system, such that the bulk densities matched those of the pure EMIMBF$_4$/ACN mixtures at \SI{1}{\bar} and \SI{340}{\kelvin}. The temperature of \SI{340}{\kelvin} was chosen to increase the ion mobility and reduce the required simulation time. 
Once correct bulk densities were achieved, the final equilibration step was performed with velocity rescaling and electrode atoms charged to constant values of $\pm$ 0.01 e for the positive and negative electrode, this step was run for several nanoseconds so that sufficient equilibrium was achieved.

For production runs, the electrodes were switched to a constant potential and the systems were run in the NVE ensemble with $\pm$ 0.5 V for the half cell potentials of the positive and negative electrodes. The length of production runs exceeded \SI{12}{\nano\second} to ensure that average electrode atom charges had reached a steady equilibrium value. 2D Ewald summation\cite{Reed2007,Gingrich2010} was used for all Coulombic interaction calculations, with a short range cut-off distance of \SI{22}{\angstrom} (half the length of the x and y cell dimensions). Lorentz-Berthelot combination mixing rules were employed for non-bonded potential interactions.

The capacitance  was calculated using $C= \frac{Q^\pm}{V^\pm} $, where $Q$ and $V$ are the electrode charge and voltage, respectively. In the simulations, $V$ is a known and constrained parameter, and $Q$ can be measured directly from the simulations.  

\section{Acknowledgments}
\label{sec:acknow}

C.H. and K.B. acknowledge the Deutsche Forschungsgemeinschaft (DFG) through the cluster of excellence ``Simulation Technology'' and the SFB 716 for financial support. K.B. acknowledges a EU COST action (CM1206) for funding his stay in Paris through a STSM grant. P.S acknowledges the support from the European Research Council under the European Union's Seventh Framework Programme (FP/2007-2013)/ERC grant Agreement n.102539 (Advanced grant, Ionaces project). G.B., X.S.Z. and R.B. acknowledge support from The Australia Research Council (ARC) under project DP130101870. M.S. acknowledges support from the French National Research Agency (Labex STORE-EX, Grant No. ANR-10-LABX-0076) and from EoCoE, a project funded by the European Union Contract No. H2020-EINFRA-2015-1-676629. We are grateful for the computing resources on OCCIGEN (CINES, French National HPC) and CURIE (TGCC, French National HPC) obtained through the project x2016096728, on the Cray XC40 (Hornet) from the SimTech Cluster of Excellence in Stuttgart, HLRS, and from the University of Queensland Research Computing Centre.

\section{Author information}
\subsection{Author contributions}
$^*$ These authors contributed equally to this work.

\bibliography{Final_Library}

\providecommand{\latin}[1]{#1}
\providecommand*\mcitethebibliography{\thebibliography}
\csname @ifundefined\endcsname{endmcitethebibliography}
  {\let\endmcitethebibliography\endthebibliography}{}
\begin{mcitethebibliography}{43}
\providecommand*\natexlab[1]{#1}
\providecommand*\mciteSetBstSublistMode[1]{}
\providecommand*\mciteSetBstMaxWidthForm[2]{}
\providecommand*\mciteBstWouldAddEndPuncttrue
  {\def\EndOfBibitem{\unskip.}}
\providecommand*\mciteBstWouldAddEndPunctfalse
  {\let\EndOfBibitem\relax}
\providecommand*\mciteSetBstMidEndSepPunct[3]{}
\providecommand*\mciteSetBstSublistLabelBeginEnd[3]{}
\providecommand*\EndOfBibitem{}
\mciteSetBstSublistMode{f}
\mciteSetBstMaxWidthForm{subitem}{(\alph{mcitesubitemcount})}
\mciteSetBstSublistLabelBeginEnd
  {\mcitemaxwidthsubitemform\space}
  {\relax}
  {\relax}

\bibitem[Salanne \latin{et~al.}(2016)Salanne, Rotenberg, Naoi, Kaneko, Taberna,
  Grey, Dunn, and Simon]{salanne2016a}
Salanne,~M.; Rotenberg,~B.; Naoi,~K.; Kaneko,~K.; Taberna,~P.-L.; Grey,~C.~P.;
  Dunn,~B.; Simon,~P. Efficient Storage Mechanisms for Building Better
  Supercapacitors. \emph{Nat. Energy} \textbf{2016}, \emph{1}, 16070\relax
\mciteBstWouldAddEndPuncttrue
\mciteSetBstMidEndSepPunct{\mcitedefaultmidpunct}
{\mcitedefaultendpunct}{\mcitedefaultseppunct}\relax
\EndOfBibitem
\bibitem[Gogotsi and Simon(2011)Gogotsi, and Simon]{Gogotsi2011}
Gogotsi,~Y.; Simon,~P. {True Performance Metrics in Electrochemical Energy
  Storage}. \emph{Science} \textbf{2011}, \emph{334}, 917--918\relax
\mciteBstWouldAddEndPuncttrue
\mciteSetBstMidEndSepPunct{\mcitedefaultmidpunct}
{\mcitedefaultendpunct}{\mcitedefaultseppunct}\relax
\EndOfBibitem
\bibitem[Faggioli \latin{et~al.}(1999)Faggioli, Rena, Danel, Andrieu, Mallant,
  and Kahlen]{Faggioli1999}
Faggioli,~E.; Rena,~P.; Danel,~V.; Andrieu,~X.; Mallant,~R.; Kahlen,~H.
  {Supercapacitors for the Energy Management of Electric Vehicles}. \emph{J.
  Power Sources} \textbf{1999}, \emph{84}, 261--269\relax
\mciteBstWouldAddEndPuncttrue
\mciteSetBstMidEndSepPunct{\mcitedefaultmidpunct}
{\mcitedefaultendpunct}{\mcitedefaultseppunct}\relax
\EndOfBibitem
\bibitem[Dunn \latin{et~al.}(2011)Dunn, Kamath, and Tarascon]{Dunn2011}
Dunn,~B.; Kamath,~H.; Tarascon,~J.-M. {Electrical Energy Storage for the Grid:
  A Battery of Choices}. \emph{Science} \textbf{2011}, \emph{334},
  928--935\relax
\mciteBstWouldAddEndPuncttrue
\mciteSetBstMidEndSepPunct{\mcitedefaultmidpunct}
{\mcitedefaultendpunct}{\mcitedefaultseppunct}\relax
\EndOfBibitem
\bibitem[Simon and Gogotsi(2008)Simon, and Gogotsi]{Simon2008a}
Simon,~P.; Gogotsi,~Y. {Materials for Electrochemical Capacitors.} \emph{Nat.
  Mater.} \textbf{2008}, \emph{7}, 845--854\relax
\mciteBstWouldAddEndPuncttrue
\mciteSetBstMidEndSepPunct{\mcitedefaultmidpunct}
{\mcitedefaultendpunct}{\mcitedefaultseppunct}\relax
\EndOfBibitem
\bibitem[Rogers \latin{et~al.}(2009)Rogers, S{\`{I}}ƒljukic, Hardacre, and
  Compton]{Rogers2009}
Rogers,~E.~I.; Sljukic,~B.; Hardacre,~C.; Compton,~R.~G.
  {Electrochemistry in Room-Temperature Ionic Liquids: Potential Windows at
  Mercury Electrodes}. \emph{J. Chem. Eng. Data} \textbf{2009}, \emph{54},
  2049--2053\relax
\mciteBstWouldAddEndPuncttrue
\mciteSetBstMidEndSepPunct{\mcitedefaultmidpunct}
{\mcitedefaultendpunct}{\mcitedefaultseppunct}\relax
\EndOfBibitem
\bibitem[Huddleston \latin{et~al.}(2001)Huddleston, Visser, Reichert, Willauer,
  Broker, and Rogers]{Huddleston2001}
Huddleston,~J.~G.; Visser,~A.~E.; Reichert,~W.~M.; Willauer,~H.~D.;
  Broker,~G.~A.; Rogers,~R.~D. {Characterization and Comparison of Hydrophilic
  and Hydrophobic Room Temperature Ionic Liquids Incorporating the Imidazolium
  Cation}. \emph{Green Chem.} \textbf{2001}, \emph{3}, 156--164\relax
\mciteBstWouldAddEndPuncttrue
\mciteSetBstMidEndSepPunct{\mcitedefaultmidpunct}
{\mcitedefaultendpunct}{\mcitedefaultseppunct}\relax
\EndOfBibitem
\bibitem[Buzzeo \latin{et~al.}(2004)Buzzeo, Evans, and Compton]{Buzzeo2004}
Buzzeo,~M.~C.; Evans,~R.~G.; Compton,~R.~G. {Non-Haloaluminate Room-Temperature
  Ionic Liquids in Electrochemistry - A Review}. \emph{ChemPhysChem}
  \textbf{2004}, \emph{5}, 1106--1120\relax
\mciteBstWouldAddEndPuncttrue
\mciteSetBstMidEndSepPunct{\mcitedefaultmidpunct}
{\mcitedefaultendpunct}{\mcitedefaultseppunct}\relax
\EndOfBibitem
\bibitem[Largeot \latin{et~al.}(2008)Largeot, Portet, Chmiola, Taberna,
  Gogotsi, and Simon]{Largeot2008}
Largeot,~C.; Portet,~C.; Chmiola,~J.; Taberna,~P.~L.; Gogotsi,~Y.; Simon,~P.
  {Relation Between the Ion Size and Pore Size for an Electric Double-Layer
  Capacitor}. \emph{J. Am. Chem. Soc.} \textbf{2008}, \emph{130},
  2730--2731\relax
\mciteBstWouldAddEndPuncttrue
\mciteSetBstMidEndSepPunct{\mcitedefaultmidpunct}
{\mcitedefaultendpunct}{\mcitedefaultseppunct}\relax
\EndOfBibitem
\bibitem[Guerfi \latin{et~al.}(2010)Guerfi, Dontigny, Charest, Petitclerc,
  Lagac{\'{e}}, Vijh, and Zaghib]{Guerfi2010}
Guerfi,~A.; Dontigny,~M.; Charest,~P.; Petitclerc,~M.; Lagac{\'{e}},~M.;
  Vijh,~A.; Zaghib,~K. {Improved Electrolytes for Li-ion Batteries: Mixtures of
  Ionic Liquid and Organic Electrolyte with Enhanced Safety and Electrochemical
  Performance}. \emph{J. Power Sources} \textbf{2010}, \emph{195},
  845--852\relax
\mciteBstWouldAddEndPuncttrue
\mciteSetBstMidEndSepPunct{\mcitedefaultmidpunct}
{\mcitedefaultendpunct}{\mcitedefaultseppunct}\relax
\EndOfBibitem
\bibitem[Richey and Elabd(2012)Richey, and Elabd]{Richey2012}
Richey,~F.~W.; Elabd,~Y.~A. {In Situ Molecular Level Measurements of Ion
  Dynamics in an Electrochemical Capacitor}. \emph{J. Phys. Chem. Lett.}
  \textbf{2012}, \emph{3}, 3297--3301\relax
\mciteBstWouldAddEndPuncttrue
\mciteSetBstMidEndSepPunct{\mcitedefaultmidpunct}
{\mcitedefaultendpunct}{\mcitedefaultseppunct}\relax
\EndOfBibitem
\bibitem[Banuelos \latin{et~al.}(2013)Banuelos, Feng, Fulvio, Li, Rother, Dai,
  Cummings, and Wesolowski]{Banuelos2013}
Banuelos,~L.~J.; Feng,~G.; Fulvio,~P.~F.; Li,~S.; Rother,~G.; Dai,~S.;
  Cummings,~P.~T.; Wesolowski,~D.~J. {Densification of Ionic Liquid Molecules
  within a Hierarchical Nanoporous Carbon Structure Revealed by Small-Angle
  Scattering and Molecular Dynamics Simulation}. \emph{Chem. Mater.}
  \textbf{2013}, \emph{26}, 1144--1153\relax
\mciteBstWouldAddEndPuncttrue
\mciteSetBstMidEndSepPunct{\mcitedefaultmidpunct}
{\mcitedefaultendpunct}{\mcitedefaultseppunct}\relax
\EndOfBibitem
\bibitem[Forse \latin{et~al.}(2015)Forse, Griffin, Merlet, Bayley, Wang, Simon,
  and Grey]{Forse2015}
Forse,~A.~C.; Griffin,~J.~M.; Merlet,~C.; Bayley,~P.~M.; Wang,~H.; Simon,~P.;
  Grey,~C.~P. {NMR Study of Ion Dynamics and Charge Storage in Ionic Liquid
  Supercapacitors}. \emph{J. Am. Chem. Soc.} \textbf{2015}, \emph{137},
  7231--7242\relax
\mciteBstWouldAddEndPuncttrue
\mciteSetBstMidEndSepPunct{\mcitedefaultmidpunct}
{\mcitedefaultendpunct}{\mcitedefaultseppunct}\relax
\EndOfBibitem
\bibitem[Fedorov and Kornyshev(2014)Fedorov, and Kornyshev]{Fedorov2014}
Fedorov,~M.~V.; Kornyshev,~A.~A. {Ionic Liquids at Electrified Interfaces}.
  \emph{Chem. Rev.} \textbf{2014}, \emph{26}, 2978--3036\relax
\mciteBstWouldAddEndPuncttrue
\mciteSetBstMidEndSepPunct{\mcitedefaultmidpunct}
{\mcitedefaultendpunct}{\mcitedefaultseppunct}\relax
\EndOfBibitem
\bibitem[Burt \latin{et~al.}(2014)Burt, Birkett, and Zhao]{Burt2014}
Burt,~R.; Birkett,~G.; Zhao,~X.~S. {A Review of Molecular Modelling of Electric
  Double Layer Capacitors.} \emph{Phys. Chem. Chem. Phys.} \textbf{2014},
  \emph{16}, 6519--6538\relax
\mciteBstWouldAddEndPuncttrue
\mciteSetBstMidEndSepPunct{\mcitedefaultmidpunct}
{\mcitedefaultendpunct}{\mcitedefaultseppunct}\relax
\EndOfBibitem
\bibitem[Feng \latin{et~al.}(2013)Feng, Li, Presser, and Cummings]{Feng2013a}
Feng,~G.; Li,~S.; Presser,~V.; Cummings,~P.~T. {Molecular Insights into Carbon
  Supercapacitors Based on Room-Temperature Ionic Liquids}. \emph{J. Phys.
  Chem. Lett.} \textbf{2013}, \emph{4}, 3367--3376\relax
\mciteBstWouldAddEndPuncttrue
\mciteSetBstMidEndSepPunct{\mcitedefaultmidpunct}
{\mcitedefaultendpunct}{\mcitedefaultseppunct}\relax
\EndOfBibitem
\bibitem[Merlet \latin{et~al.}(2013)Merlet, P{\'{e}}an, Rotenberg, Madden,
  Simon, and Salanne]{Merlet2013f}
Merlet,~C.; P{\'{e}}an,~C.; Rotenberg,~B.; Madden,~P.~A.; Simon,~P.;
  Salanne,~M. {Simulating Supercapacitors: Can We Model Electrodes as Constant
  Charge Surfaces?} \emph{J. Phys. Chem. Lett.} \textbf{2013}, \emph{4},
  264--268\relax
\mciteBstWouldAddEndPuncttrue
\mciteSetBstMidEndSepPunct{\mcitedefaultmidpunct}
{\mcitedefaultendpunct}{\mcitedefaultseppunct}\relax
\EndOfBibitem
\bibitem[Vatamanu \latin{et~al.}(2013)Vatamanu, Hu, Bedrov, Perez, and
  Gogotsi]{Vatamanu2013}
Vatamanu,~J.; Hu,~Z.; Bedrov,~D.; Perez,~C.; Gogotsi,~Y. {Increasing Energy
  Storage in Electrochemical Capacitors with Ionic Liquid Electrolytes and
  Nanostructured Carbon Electrodes}. \emph{J. Phys. Chem. Lett.} \textbf{2013},
  \emph{4}, 2829--2837\relax
\mciteBstWouldAddEndPuncttrue
\mciteSetBstMidEndSepPunct{\mcitedefaultmidpunct}
{\mcitedefaultendpunct}{\mcitedefaultseppunct}\relax
\EndOfBibitem
\bibitem[He \latin{et~al.}(2015)He, Huang, Sumpter, Kornyshev, and
  Qiao]{he2015aa}
He,~Y.; Huang,~J.; Sumpter,~B.~G.; Kornyshev,~A.~A.; Qiao,~R. Dynamic Charge
  Storage in Ionic Liquids-filled Nanopores: Insight from a Computational
  Cyclic Voltammetry Study. \emph{J. Phys. Chem. Lett.} \textbf{2015},
  \emph{6}, 22--30\relax
\mciteBstWouldAddEndPuncttrue
\mciteSetBstMidEndSepPunct{\mcitedefaultmidpunct}
{\mcitedefaultendpunct}{\mcitedefaultseppunct}\relax
\EndOfBibitem
\bibitem[Feng \latin{et~al.}(2011)Feng, Huang, Sumpter, Meunier, and
  Qiao]{Feng2011b}
Feng,~G.; Huang,~J.; Sumpter,~B.~G.; Meunier,~V.; Qiao,~R. {A "Counter-Charge
  Layer in Generalized Solvents" Framework for Electrical Double Layers in Neat
  and Hybrid Ionic Liquid Electrolytes.} \emph{Phys. Chem. Chem. Phys.}
  \textbf{2011}, \emph{13}, 14723--14734\relax
\mciteBstWouldAddEndPuncttrue
\mciteSetBstMidEndSepPunct{\mcitedefaultmidpunct}
{\mcitedefaultendpunct}{\mcitedefaultseppunct}\relax
\EndOfBibitem
\bibitem[Uralcan \latin{et~al.}(2016)Uralcan, Aksay, Debenedetti, and
  Limmer]{Uralcan2016}
Uralcan,~B.; Aksay,~I.~A.; Debenedetti,~P.~G.; Limmer,~D.~T. {Concentration
  Fluctuations and Capacitive Response in Dense Ionic Solutions}. \emph{J.
  Phys. Chem. Lett.} \textbf{2016}, \emph{7}, 2333--2338\relax
\mciteBstWouldAddEndPuncttrue
\mciteSetBstMidEndSepPunct{\mcitedefaultmidpunct}
{\mcitedefaultendpunct}{\mcitedefaultseppunct}\relax
\EndOfBibitem
\bibitem[Lee \latin{et~al.}(2016)Lee, and
  Perkin]{lee2016b}
Lee,~A.~A.; Perkin,~S. {Ion-Image
  Interactions and Phase Transition at Electrolyte-Metal Interfaces}. \emph{J.
  Phys. Chem. Lett.} \textbf{2016}, \emph{7}, 2753--2757\relax
\mciteBstWouldAddEndPuncttrue
\mciteSetBstMidEndSepPunct{\mcitedefaultmidpunct}
{\mcitedefaultendpunct}{\mcitedefaultseppunct}\relax
\EndOfBibitem
\bibitem[Reed \latin{et~al.}(2007)Reed, Lanning, and Madden]{Reed2007}
Reed,~S.~K.; Lanning,~O.~J.; Madden,~P.~A. {Electrochemical Interface Between
  an Ionic Liquid and a Model Metallic Electrode}. \emph{J. Chem. Phys.}
  \textbf{2007}, \emph{126}, 084704\relax
\mciteBstWouldAddEndPuncttrue
\mciteSetBstMidEndSepPunct{\mcitedefaultmidpunct}
{\mcitedefaultendpunct}{\mcitedefaultseppunct}\relax
\EndOfBibitem
\bibitem[Pounds \latin{et~al.}(2009)Pounds, Tazi, Salanne, and
  Madden]{Pounds2009}
Pounds,~M.; Tazi,~S.; Salanne,~M.; Madden,~P.~A. {Ion Adsorption at a Metallic
  Electrode: an Ab Initio Based Simulation Study.} \emph{J. Phys.: Condens.
  Matter} \textbf{2009}, \emph{21}, 424109\relax
\mciteBstWouldAddEndPuncttrue
\mciteSetBstMidEndSepPunct{\mcitedefaultmidpunct}
{\mcitedefaultendpunct}{\mcitedefaultseppunct}\relax
\EndOfBibitem
\bibitem[Kondrat \latin{et~al.}(2012)Kondrat, Perez, Presser, Gogotsi, and
 Kornyshev]{Kondrat2012}
Kondrat,~S.; P\'erez,~C.~R.; Presser,~V.; Gogotsi,~Y;
  Kornyshev,~A.~A. {Effect of pore size and its dispersity on the energy storage in nanoporous supercapacitors.} \emph{Eng. Env. Sci.} \textbf{2012}, \emph{5},
  6474--6479\relax
\mciteBstWouldAddEndPuncttrue
\mciteSetBstMidEndSepPunct{\mcitedefaultmidpunct}
{\mcitedefaultendpunct}{\mcitedefaultseppunct}\relax
\EndOfBibitem
\bibitem[Palmer \latin{et~al.}(2010)Palmer, Llobet, Yeon, Fischer, Shi,
  Gogotsi, and Gubbins]{Palmer2010}
Palmer,~J.~C.; Llobet,~A.; Yeon,~S.~H.; Fischer,~J.~E.; Shi,~Y.; Gogotsi,~Y.;
  Gubbins,~K.~E. {Modeling the Structural Evolution of Carbide-Derived Carbons
  Using Quenched Molecular Dynamics}. \emph{Carbon} \textbf{2010}, \emph{48},
  1116--1123\relax
\mciteBstWouldAddEndPuncttrue
\mciteSetBstMidEndSepPunct{\mcitedefaultmidpunct}
{\mcitedefaultendpunct}{\mcitedefaultseppunct}\relax
\EndOfBibitem
\bibitem[Vatamanu \latin{et~al.}(2014)Vatamanu, Xing, Li, and
  Bedrov]{Vatamanu2014}
Vatamanu,~J.; Xing,~L.; Li,~W.; Bedrov,~D. {Influence of Temperature on the
  Capacitance of Ionic Liquid Electrolytes on Charged Surfaces.} \emph{Phys.
  Chem. Chem. Phys.} \textbf{2014}, \emph{16}, 5174--5182\relax
\mciteBstWouldAddEndPuncttrue
\mciteSetBstMidEndSepPunct{\mcitedefaultmidpunct}
{\mcitedefaultendpunct}{\mcitedefaultseppunct}\relax
\EndOfBibitem
\bibitem[Kornyshev \latin{et~al.}(2014)Kornyshev, Luque, and
  Schmickler]{Kornyshev2014}
Kornyshev,~A.~A.; Luque,~N.~B.; Schmickler,~W. {Differential Capacitance of
  Ionic Liquid Interface with Graphite: The Story of two Double Layers}.
  \emph{J. Solid State Electrochem.} \textbf{2014}, \emph{18}, 1345--1349\relax
\mciteBstWouldAddEndPuncttrue
\mciteSetBstMidEndSepPunct{\mcitedefaultmidpunct}
{\mcitedefaultendpunct}{\mcitedefaultseppunct}\relax
\EndOfBibitem
\bibitem[P{\'{e}}an \latin{et~al.}(2015)P{\'{e}}an, Daffos, Rotenberg, Levitz,
  Haefele, Taberna, Simon, and Salanne]{pean2015b}
P{\'{e}}an,~C.; Daffos,~B.; Rotenberg,~B.; Levitz,~P.; Haefele,~M.;
  Taberna,~P.~L.; Simon,~P.; Salanne,~M. {Confinement, Desolvation, and
  Electrosorption Effects on the Diffusion of Ions in Nanoporous Carbon
  Electrodes}. \emph{J. Am. Chem. Soc.} \textbf{2015}, \emph{137},
  12627--12632\relax
\mciteBstWouldAddEndPuncttrue
\mciteSetBstMidEndSepPunct{\mcitedefaultmidpunct}
{\mcitedefaultendpunct}{\mcitedefaultseppunct}\relax
\EndOfBibitem
\bibitem[Forse \latin{et~al.}(2016)Forse, Merlet, Griffin, and Grey]{Forse2016}
Forse,~A.~C.; Merlet,~C.; Griffin,~J.~M.; Grey,~C.~P. {New Perspectives on the
  Charging Mechanisms of Supercapacitors}. \emph{J. Am. Chem. Soc.}
  \textbf{2016}, \emph{138}, 5731--5744\relax
\mciteBstWouldAddEndPuncttrue
\mciteSetBstMidEndSepPunct{\mcitedefaultmidpunct}
{\mcitedefaultendpunct}{\mcitedefaultseppunct}\relax
\EndOfBibitem
\bibitem[Kondrat and Kornyshev(2011)Kondrat, and Kornyshev]{Kondrat2011c}
Kondrat,~S.; Kornyshev,~A. {Superionic State in Double-Layer Capacitors with
  Nanoporous Electrodes.} \emph{J. Phys.: Condens. Matter} \textbf{2011},
  \emph{23}, 022201\relax
\mciteBstWouldAddEndPuncttrue
\mciteSetBstMidEndSepPunct{\mcitedefaultmidpunct}
{\mcitedefaultendpunct}{\mcitedefaultseppunct}\relax
\EndOfBibitem
\bibitem[Merlet \latin{et~al.}(2012)Merlet, Rotenberg, Madden, Taberna, Simon,
  Gogotsi, and Salanne]{Merlet2012e}
Merlet,~C.; Rotenberg,~B.; Madden,~P.~A.; Taberna,~P.-L.; Simon,~P.;
  Gogotsi,~Y.; Salanne,~M. {On the Molecular Origin of Supercapacitance in
  Nanoporous Carbon Electrodes}. \emph{Nat. Mater.} \textbf{2012}, \emph{11},
  306--310\relax
\mciteBstWouldAddEndPuncttrue
\mciteSetBstMidEndSepPunct{\mcitedefaultmidpunct}
{\mcitedefaultendpunct}{\mcitedefaultseppunct}\relax
\EndOfBibitem
\bibitem[Merlet \latin{et~al.}(2012)Merlet, Salanne, and
  Rotenberg]{Merlet2012g}
Merlet,~C.; Salanne,~M.; Rotenberg,~B. {New Coarse-Grained Models of
  Imidazolium Ionic Liquids for Bulk and Interfacial Molecular Simulations}.
  \emph{J. Phys. Chem. C} \textbf{2012}, \emph{116}, 7687--7693\relax
\mciteBstWouldAddEndPuncttrue
\mciteSetBstMidEndSepPunct{\mcitedefaultmidpunct}
{\mcitedefaultendpunct}{\mcitedefaultseppunct}\relax
\EndOfBibitem
\bibitem[B\'eguin \latin{et~al.}(2014)B\'eguin, Presser, Balducci, and
  Frackowiak]{beguin2014a}
B\'eguin,~F.; Presser,~V.; Balducci,~A.; Frackowiak,~E. Carbons and
  Electrolytes for Advanced Supercapacitors. \emph{Adv. Mater.} \textbf{2014},
  \emph{26}, 2219--2251\relax
\mciteBstWouldAddEndPuncttrue
\mciteSetBstMidEndSepPunct{\mcitedefaultmidpunct}
{\mcitedefaultendpunct}{\mcitedefaultseppunct}\relax
\EndOfBibitem
\bibitem[Krause and Balducci(2011)Krause, and Balducci]{krause2011a}
Krause,~A.; Balducci,~A. High Voltage Electrochemical Double Layer Capacitor
  Containing Mixtures of Ionic Liquids and Organic Carbonate as Electrolytes.
  \emph{Electrochem. Commun.} \textbf{2011}, \emph{13}, 814--817\relax
\mciteBstWouldAddEndPuncttrue
\mciteSetBstMidEndSepPunct{\mcitedefaultmidpunct}
{\mcitedefaultendpunct}{\mcitedefaultseppunct}\relax
\EndOfBibitem
\bibitem[Pohlmann \latin{et~al.}(2015)Pohlmann, Olyschlager, Goodrich, Vicente,
  Jacquemin, and Balducci]{pohlmann2015a}
Pohlmann,~S.; Olyschlager,~T.; Goodrich,~P.; Vicente,~J.~A.; Jacquemin,~J.;
  Balducci,~A. Mixtures of Azepanium Based Ionic Liquids and Propylene
  Carbonate as High Voltage Electrolytes for Supercapacitors.
  \emph{Electrochim. Acta} \textbf{2015}, \emph{153}, 426--432\relax
\mciteBstWouldAddEndPuncttrue
\mciteSetBstMidEndSepPunct{\mcitedefaultmidpunct}
{\mcitedefaultendpunct}{\mcitedefaultseppunct}\relax
\EndOfBibitem
\bibitem[Lin \latin{et~al.}(2011)Lin, Taberna, Fantini, Presser, Perez,
  Malbosc, Rupesinghe, Teo, Gogotsi, and Simon]{lin2011a}
Lin,~R.~Y.; Taberna,~P.-L.; Fantini,~S.; Presser,~V.; Perez,~C.~R.;
  Malbosc,~F.; Rupesinghe,~N.~L.; Teo,~K. B.~K.; Gogotsi,~Y.; Simon,~P.
  Capacitive Energy Storage from -50 to 100 degrees C Using an Ionic Liquid
  Electrolyte. \emph{J. Phys. Chem. Lett.} \textbf{2011}, \emph{2},
  2396--2401\relax
\mciteBstWouldAddEndPuncttrue
\mciteSetBstMidEndSepPunct{\mcitedefaultmidpunct}
{\mcitedefaultendpunct}{\mcitedefaultseppunct}\relax
\EndOfBibitem
\bibitem[Sch\"utter \latin{et~al.}(2015)Sch\"utter, Husch, Korth, and
  Balducci]{schutter2015a}
Sch\"utter,~C.; Husch,~T.; Korth,~M.; Balducci,~A. Toward New Solvents for
  {EDLC}s: From Computational Screening to Electrochemical Validation. \emph{J.
  Phys. Chem. C} \textbf{2015}, \emph{119}, 13413--13424\relax
\mciteBstWouldAddEndPuncttrue
\mciteSetBstMidEndSepPunct{\mcitedefaultmidpunct}
{\mcitedefaultendpunct}{\mcitedefaultseppunct}\relax
\EndOfBibitem
\bibitem[Chmiola \latin{et~al.}(2006)Chmiola, Yushin, Gogotsi, Portet, Simon,
  and Taberna]{chmiola2006a}
Chmiola,~J.; Yushin,~G.; Gogotsi,~Y.; Portet,~C.; Simon,~P.; Taberna,~P.~L.
  {Anomalous Increase in Carbon Capacitance at Pore Sizes less than 1
  nanometer.} \emph{Science} \textbf{2006}, \emph{313}, 1760--1763\relax
\mciteBstWouldAddEndPuncttrue
\mciteSetBstMidEndSepPunct{\mcitedefaultmidpunct}
{\mcitedefaultendpunct}{\mcitedefaultseppunct}\relax
\EndOfBibitem
\bibitem[Portet \latin{et~al.}(2005)Portet, Taberna, Simon, Flahaut, and
  Laberty-Robert]{Portet2005}
Portet,~C.; Taberna,~P.~L.; Simon,~P.; Flahaut,~E.; Laberty-Robert,~C. {High
  Power Density Electrodes for Carbon Supercapacitor Applications}.
  \emph{Electrochim. Acta} \textbf{2005}, \emph{50}, 4174--4181\relax
\mciteBstWouldAddEndPuncttrue
\mciteSetBstMidEndSepPunct{\mcitedefaultmidpunct}
{\mcitedefaultendpunct}{\mcitedefaultseppunct}\relax
\EndOfBibitem
\bibitem[Edwards \latin{et~al.}(1984)Edwards, Madden, and
  McDonald]{Edwards1984}
Edwards,~D. M.~F.; Madden,~P.~A.; McDonald,~I.~R. {A Computer Simulation Study
  of the Dielectric Properties of a Model of Methyl Cyanide}. \emph{Mol. Phys.}
  \textbf{1984}, \emph{51}, 1141--1161\relax
\mciteBstWouldAddEndPuncttrue
\mciteSetBstMidEndSepPunct{\mcitedefaultmidpunct}
{\mcitedefaultendpunct}{\mcitedefaultseppunct}\relax
\EndOfBibitem
\bibitem[Cole and Klein(1983)Cole, and Klein]{cole1983}
Cole,~M.~W.; Klein,~J.~R. {The Interaction Between Noble Gases and the Basal
  Plane Surface of Graphite}. \emph{Surf. Sci.} \textbf{1983}, \emph{124},
  547--554\relax
\mciteBstWouldAddEndPuncttrue
\mciteSetBstMidEndSepPunct{\mcitedefaultmidpunct}
{\mcitedefaultendpunct}{\mcitedefaultseppunct}\relax
\EndOfBibitem
\bibitem[Hess \latin{et~al.}(2008)Hess, Kutzner, {van der Spoel}, and
  Lindahl]{hess08a}
Hess,~B.; Kutzner,~C.; {van der Spoel},~D.; Lindahl,~E. {GROMACS} 4: Algorithms
  for Highly Efficient, Load-Balanced, and Scalable Molecular Simulation.
  \emph{J. Chem. Theor Comput.} \textbf{2008}, \emph{4}, 435--447\relax
\mciteBstWouldAddEndPuncttrue
\mciteSetBstMidEndSepPunct{\mcitedefaultmidpunct}
{\mcitedefaultendpunct}{\mcitedefaultseppunct}\relax
\EndOfBibitem
\bibitem[Gingrich and Wilson(2010)Gingrich, and Wilson]{Gingrich2010}
Gingrich,~T.~R.; Wilson,~M. {On the Ewald Summation of Gaussian Charges for the
  Simulation of Metallic Surfaces}. \emph{Chem. Phys. Lett.} \textbf{2010},
  \emph{500}, 178--183\relax
\mciteBstWouldAddEndPuncttrue
\mciteSetBstMidEndSepPunct{\mcitedefaultmidpunct}
{\mcitedefaultendpunct}{\mcitedefaultseppunct}\relax
\EndOfBibitem
\end{mcitethebibliography}

\end{document}